\documentclass[conference]{IEEEtran}
\IEEEoverridecommandlockouts
\usepackage{amsmath,amssymb,amsfonts}
\usepackage{algorithmic}
\usepackage{graphicx}
\usepackage{textcomp}
\usepackage{xcolor}
\usepackage[nolist]{acronym}

\usepackage[numbers, sort&compress]{natbib}
\setcitestyle{open={},close={}} 
 
\usepackage{makecell}
\usepackage{booktabs}
\usepackage[para,online,flushleft]{threeparttable}

\usepackage{hyperref}
\usepackage{lipsum}
\usepackage{dblfloatfix}
\usepackage{float}
\usepackage{footmisc}
\usepackage{listings}

\definecolor{really-light-gray}{gray}{0.95}

\def\BibTeX{{\rm B\kern-.05em{\sc i\kern-.025em b}\kern-.08em T\kern-.1667em\lower.7ex\hbox{E}\kern-.125emX}}

\def\BibTeX{{\rm B\kern-.05em{\sc i\kern-.025em b}\kern-.08em
    T\kern-.1667em\lower.7ex\hbox{E}\kern-.125emX}}
\usepackage{tikz}

\makeatletter
\newcommand{\setword}[2]{%
  \phantomsection
  #1\def\@currentlabel{\unexpanded{#1}}\label{#2}%
}
\makeatother
\begin{document}
\bstctlcite{IEEEexample:BSTcontrol} 
\title{A Formal Model for Artificial Intelligence Applications in Automation Systems}
\author{
\IEEEauthorblockN{Marvin Schieseck, Philip Topalis, Lasse Reinpold, Felix Gehlhoff}
\IEEEauthorblockA{
    \textit{Institute of Automation Technology} \\
    \textit{Helmut Schmidt University}\\
    Hamburg, Germany \\
    marvin.schieseck@hsu-hh.de
    }
\and
\IEEEauthorblockN{Alexander Fay}
\IEEEauthorblockA{\textit{Chair of Automation} \\
\textit{Ruhr University}\\
Bochum, Germany \\
alexander.fay@rub.de}
}

\maketitle
\begin{abstract} 
The integration of Artificial Intelligence (AI) into automation systems has the potential to enhance efficiency and to address currently unsolved existing technical challenges. However, the industry-wide adoption of AI is hindered by the lack of standardized documentation for the complex compositions of automation systems, AI software, production hardware, and their interdependencies. This paper proposes a formal model using standards and ontologies to provide clear and structured documentation of AI applications in automation systems. The proposed information model for artificial intelligence in automation systems (AIAS) utilizes ontology design patterns to map and link various aspects of automation systems and AI software. Validated through a practical example, the model demonstrates its effectiveness in improving documentation practices and aiding the sustainable implementation of AI in industrial settings.
\end{abstract}
 
\begin{IEEEkeywords}
Artificial Intelligence, Modeling, Semantic Web, Web Ontology Language, OWL
\end{IEEEkeywords}
\section{Introduction} \label{02_introduction}
The digitization and automation of technical plants, manufacturing processes, and products are continually increasing, driven by modern information technologies. To stay competitive in the global market, companies are forced to adapt to these technologies such as \ac{ai}. \cite{OECD-2019-a}

As a result, the integration of \ac{ai} applications into automation systems has been increasingly driven forward in recent years \cite{wef2023}. Numerous use cases have already been identified in which the integration of AI has enabled new solutions or increased the efficiency of previous solutions \cite{Peres-2020}. Examples can be found in areas like maintenance, quality control, and planning \cite{ISO-TR-24030}. Despite these promising use cases, the adoption of AI across industries remains relatively low \cite{wef2023}, \cite{merkel-kiss2023}. 

AI applications are often developed as part of research projects and cannot be economically used in practice. This is partly because some aspects of AI-software, such as data quality and management, computing resources management, trustworthiness, and software documentation are particularly challenging compared to conventional automation software and hinder economic integration and operation. \cite{merkel-kiss2023, hoffmann-2021, paleyes2022, Sculley-2015}

While some of these challenging aspects, such as data quality or trustworthiness, have received extensive research attention in the past years, the software documentation of \ac{ai} application and its impact to the overall system in various scenarios is often disregarded \cite{paleyes2022}, \cite{Sculley-2015}. 

At the same time, there is a global trend towards the creation of new legal regulations for AI, as exemplified by the EU AI Act \cite{eu-ai-act-2024}, which mandates a comprehensive analysis of AI application risks and recommends their documentation. As these regulations continue to evolve, clear and structured documentation of AI applications will become increasingly important. Therefore, it is reasonable to assume that a semantically unambiguous and clearly structured documentation of AI applications will become even more important in the future. 

Additional difficulties in documentation arise from the inherent complexity of automation systems and AI systems, as both systems usually consist of several hardware and software components that are highly interconnected and interdependent \cite{Filho-2022}. Moreover, automation systems often have long lifespans, spanning decades. When individual technical components within these systems are updated or replaced, comprehensive documentation is required to determine how changes, such as sensor or controller replacements, affect the AI application. Consequently, a detailed  documentation becomes necessary for deploying and operating AI applications in automation systems.

However, there is currently no suitable approach to document AI applications in automation systems that captures information about automation system components, AI software components, and the underlying technical process as well as their interdependencies and relationships. In particular, there is a lack of formal models to describe AI applications within automation systems \cite{acatech-2021}. In response, this paper proposes a new formal model for \acfi{aias} based on standards and utilizing ontologies. The primary contributions of this paper include:

\begin{itemize}
    \item A discussion of existing modeling approaches and formal models in Section \ref{sec:related_works}.
    \item A presentation of the proposed formal model AIAS based on ontologies in Section \ref{sec:method}.
    \item An application example for AIAS based on an industrial use case in Section \ref{sec:validation}.
\end{itemize}
\section{Analysis of Requirements} \label{requirements}
Software for AI applications has additional complexity compared to traditional software. These include aspects such as unclear system boundaries, undeclared data dependencies, configuration issues, and changes in the external environment, just to name a few. Notably, the code for AI models and their training often comprises only a few lines of program code, representing a small portion of the total code. The majority of an AI application respectively its program code deals with tasks such as automation, testing, resource management, process management, deployment as well as data collection, storage, transfer and verification. \cite{Sculley-2015}


Furthermore, changes that cause future data to differ from historical data can have a significant impact on the entire AI application. Therefore, replacing, updating, or calibrating hardware components such as sensors or actuators may negatively affect the overall AI application, even if it improves the data quality overall. \cite{Sculley-2015}

Moreover, most AI applications in automation systems are decentralized and interconnected systems, characterized by a multitude of software and hardware components distributed throughout the entire automation system \cite{Filho-2022}.

\subsection{Requirements for the Information Model}
Based on the challenges and aspects presented, the requirements (R) for a formal model for AI applications in automation systems are derived in the following:

\textbf{R1) Description of Interdependencies}: 
The approach must capture information and interdependencies among automation system components, AI software components, and the technical process. This is essential because AI applications in automation systems are typically distributed across several components within the technical system and therefore have to cover various dependencies \cite{Filho-2022}.

\textbf{R2) Semantic Clarity and Standardization}: 
The approach must ensure semantic clarity to enable various experts from different domains to understand, document, and communicate information in a clear manner with a clear meaning. To achieve this, the content should adhere to standards, as they are developed by groups of experts and represent a semantic consensus within specific communities \cite{hildebrandt2020}. Another advantage is that they are independent of vendors.

\textbf{R3) Formalized Representation}: 
The approach must be able to represent the information in a formalized way to ensure reusability and minimize ambiguity \cite{acatech-2021}. In addition, such formalization in combination with a vendor-independent exchange format makes the approach machine-readable and facilitates integration into the diverse tool landscape of automation systems and software engineering.

\textbf{R4) Expandability and Adaptability}:  
The approach must be expandable and adaptable to accommodate changes in standards, regulations and legislations, particularly in the rapidly evolving field of AI \cite{acatech-2021}. As AI regulations vary across countries and undergo frequent updates, the approach will require numerous adaptations. In contrast, the parts of the approach describing information about automation systems may remain stable over decades.

In accordance with these requirements, we have formulated questions together with partners from industry that the approach must be able to answer. In Tab. \ref{table:ComQuestions} a small excerpt of these questions is shown as an example.

\begin{table}[!h]
\caption{Exemplary Questions}
\begin{center}
\begin{threeparttable}
\begin{tabular}{c c}
    \toprule
    No. & \textbf{Example Questions:} \\ \midrule
    1  &  \makecell[l]{Where is the model trained?} \\
    2  &  \makecell[l]{What communication path does the model use?} \\
    3  &  \makecell[l]{Where is the production data recorded?} \\
    4  &  \makecell[l]{Which kind of task does the model solve?} \\ \bottomrule 
\end{tabular}
\begin{tablenotes}
\end{tablenotes}
\end{threeparttable}
\end{center}
\label{table:ComQuestions}
\end{table}


\section{Related Works} \label{sec:related_works}
We identified three areas that address relevant research: 1.) \textit{datasheets and cards}, 2.) \textit{ontologies}, and 3.) \textit{graphical modeling}. Some of the most relevant works in the respective areas are briefly presented in the following.

\subsection{Datasheets and Cards}
One area of research focuses on creating \textit{sheets} and \textit{cards}, which are short documents with predefined categories to capture relevant information about the AI application. In order to obtain the relevant information, the authors use a number of predefined questions that must be answered and written down by the developers of the AI application.

For example, \citet{gebru2018} propose an approach called \textit{Datasheets for Datasets} to reflect and document the creation process, distribution, maintenance, assumptions, and risks of a dataset. Inspired by Datasheets for Datasets, the authors \citet{mitchell2019} propose a framework called \textit{Model Cards} to encourage transparent model reporting. Model Cards are short documents for trained machine learning (ML) models to provide benchmarks of the model performance in different conditions. \citet{arnold2019} propose \textit{FactSheets} to increase trust in AI services by documenting various characteristics of AI services, including whether the data used to develop the services are accompanied by datasheets. Similarly, \citet{lavin2022} propose a process to ensure the documentation of robust, reliable and responsible machine learning systems through so-called \textit{Technology Readiness Level Cards} (TRL cards).

\subsection{Ontologies}
The second area of research focuses on documenting AI applications in a formal way by using ontologies instead of sheets and cards. This research area can be divided into two major fields \cite{sinha2022dm}, \cite{sinha2021ml}. One field focuses on ontologies to support the creation and documentation of the underlying development process, as summarized by \citet{sinha2022dm}. For example, \citet{panov2013} propose an ontology called \textit{OntoDM-KDD} for representing the knowledge discovery process based on the \textit{Cross Industry Standard Process for Data Mining} (CRISP-DM). Another ontology, the \textit{Data Mining Optimization Ontology} (DMOP) by \citet{keet2015}, simplifies decision-making and optimizes the performance of data mining processes. 

The second field focuses on ontologies that support the documentation of models and algorithms, especially for machine learning (ML), as conducted by \citet{sinha2021ml}. For instance, the purpose of the \textit{Exposé} ontology proposed by \citet{vanschoren2010} is to model some data mining experiments. \citet{esteves2015} propose an ontology called \textit{MEX Vocabulary}, which purpose is to describe ML experiments. \citet{publio2018} propose the \textit{ML-Schema} ontology for representing and interchanging information on ML algorithms, datasets, and experiments. These ontology-based approaches for ML serve as the basis for large community platforms such as OpenML\footnote{\url{https://www.openml.org/}} for exchanging information about ML approaches \cite{vanschoren2014}.

\subsection{Graphical Modeling}
The third area of research focuses on using graphical modeling. A graphical model can be used to represent the core of a problem in a semi-formal and understandable way for different stakeholders and experts. An approach specifically for the graphical representation of AI systems was presented by \citet{kaymakci2021}, centered on presenting relations between data sources, data sinks, and data-processing components. On this basis, a graphical modeling language was proposed by \citet{schieseck2023} which can represent AI systems within automation systems. Using the systems engineering principles outlined by \citet{haberfellner2019}, their modeling language categorizes the entire system into three primary element types: 1.) system components, 2.) system functions, and 3.) system relations. These relations, encompassing communication, assignment, and product flow, establish the interconnections among the system components (e.g., sensor, actuator, controller, cloud) and system functions (e.g., train, record, store, inference). The authors provide a set of symbols and a syntax to represent these element types. Additionally, they offer a metamodel\footnote{\url{https://github.com/schiesem/GML-AIAAS} \label{fn:metamodel}} to further formalize their approach.

\subsection{Discussion of Related Works}
After presenting the relevant areas and related works, a discussion is conducted based on the presented requirements.

The \textit{datasheets and cards} only allow a description of the AI components and the data. Documenting the technical system, the technical process as well its interdependencies is not included and not intended by the authors (R1). Despite being organized into predefined categories and questions, responses are primarily given in free-text format (R3). Consequently, formality and accuracy depend on the respondent's integrity and experience \cite{mitchell2019}. The cards and sheets do not adhere to vocabularies and terms of standards or regulations (R2). However, their straightforward structure facilitates easy expansion and adaptation to various use cases or domains (R4).

The approaches based on \textit{ontologies} are, by their very nature, formalized (R3). Most of the approaches are using the \textit{Web Ontology Language}\footnote{\url{https://www.w3.org/OWL/}} (OWL) as a vendor-independent format (R3). Neither the ontologies focussing on data mining processes nor those focussing on ML algorithms and models allow the consideration of the technical process, the technical system, the AI components and their interdependencies (R1). Furthermore, none of these ontologies do not adhere or reference to standards or regulations (R2). In addition, most of the ontologies are not semantically consistent or aligned with each other. Most of the presented ontologies are large, monolithic blocks consisting of many classes and relationships. While theoretically easy to expand and adapt, their large, monolithic structures present practical challenges (R4). Other ontologies that focus on the description of technical processes or resources, such as the \textit{Process Specification Language} (PSL) ontology \cite{gruninger2004} or the \textit{Semantic Sensor Network} (SSN)\footref{fn:ssn} ontology, do not take into account the AI components (R1).

Besides the approaches using sheets and cards or ontologies, also approaches using \textit{graphical modeling} were presented. In its core, the approach of \citet{schieseck2023} allows the representation of interdependencies between technical system, technical process and AI functions (R1) through a defined symbolism and syntax. By additionally using the provided metamodel, the symbolism can be transformed and formalized in a vendor-neutral format (R3), e.g. XML or JSON. Notably, this approach does not adhere to or references standards or regulations (R2). Due to the simple structure of the metamodel, the content can be easily expanded and adapted (R4).

In summary, none of the existing approaches fulfills the specified requirements to address the identified problem. 

\section{Formal Artificial Intelligence Application Description}  \label{sec:method}
In this section, we introduce the Information Model for \textit{Artificial Intelligence in Automation Systems} (AIAS)\footnote{\url{https://github.com/schiesem/aias-information-model} \label{fn:odps}}. This information model is built using ontologies, as these offer a way to create semantically enriched and formalized information models. Furthermore, the use of ontologies in combination with formal description languages of the semantic web, such as OWL, enables utilizing technologies of the semantic web like reasoning, querying, or applying rules. To ensure that the information model can be easily expanded and adapted, it should not be a single, monolithic ontology. Instead, the information model is composed of small and independent ontologies known as \textit{Ontology Design Patterns} (ODPs) \cite{gangemi2009}. These ODPs are interconnected at the top-level through an alignment ontology, serving as the overarching structure of the information model. To further enhance the semantics, each ODP should be based on a standard \cite{hildebrandt2020}. We categorize the used ODPs into two categories to describe various aspects:

\begin{itemize}
    \item \textbf{Technical System and Technical Process:} This category includes ODPs that contain terms and relationships to describe the technical system structure and the underlying technical processes.
    \item \textbf{Artificial Intelligence and Data}: This category consists of ODPs containing terms and relationships to describe the AI components and functionalities.
\end{itemize}
 
We have developed multiple ODPs for each category\footref{fn:odps}. Subsequently, all categories with their respective ODPs are elaborated in more detail in the following.

\subsection{Description of the Technical System and Process}
The first requirement (R1) underscores the necessity of describing the technical system aspects, including the technical components and their underlying processes. The concepts outlined in VDI 3682 \cite{VDI-3682} provide a standardized framework for describing various technical processes and the associated system structure. According to VDI 3682, a process involves the transformation of inputs into outputs, executed by a process operator assigned to a technical resource. This technical resource then carries out the process. In a broader sense, a technical resource can be viewed as capable of implementing specific functions (e.g., processes) and serves as an abstract description of the system hardware. However, it is necessary to complement the VDI 3682 ODP by two technical aspects, which are not covered in detail by VDI 3682 itself. 

Firstly, VDI 3682 lacks specificity regarding the different types of technical resources. The \textit{ECLASS}\footnote{\url{https://eclass.eu/}} specification addresses this gap by semantically defining a wide array of classes and individuals of technical resources, including sensors, actuators, controllers, personal computers, as well as broader entities such as data centers or cloud services. An alternative to \textit{ECLASS} is the \textit{United Nations Standard Products and Services Code}\footnote{\url{https://www.unspsc.org/}} (UNSPSC). In addition, the SSN \footnote{\url{https://www.w3.org/TR/vocab-ssn/}\label{fn:ssn}} ontologie provides an alternative for specifying sensors and actuators, offering a comprehensive vocabulary if needed.

\begin{figure}[!h]
    \centering
    \includegraphics[]{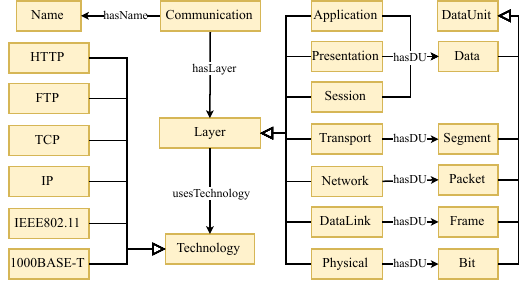}
    \caption{Core concepts and relations of the ISO 7489 ODP.} 
    \vspace*{-3mm}
    \label{fig:iso7489}
\end{figure}

Secondly, the VDI 3682 does not specify the communication between different technical resources. A widely used standard to describe such communications between technical systems is provided by ISO 7498 \cite{ISO-7498}. It is based on a seven layer communication model, called \textit{Open Systems Interconnection} (OSI) model. Using this OSI model, a communication technology can be specified across the layers, from the physical level right up to the application level. The ODP we have developed for the ISO 7489 is shown in Figure \ref{fig:iso7489}.

\subsection{Description of the Artificial Intelligence and Data}
In addition to the technical aspects, the first requirement (R1) emphasizes the necessity of describing the AI components and the utilized data. ISO 22989 \cite{ISO-22989} offers appropriate concepts and terminology for AI. This standard is structured around three overarching aspects: 1.) components and functions, 2.) algorithms, and 3.) data. Given the complexity and interrelated nature of the ISO 22989, it is useful to divide it into different viewpoints for each aspect. 

\begin{figure}[!h]
    \centering
    \includegraphics[]{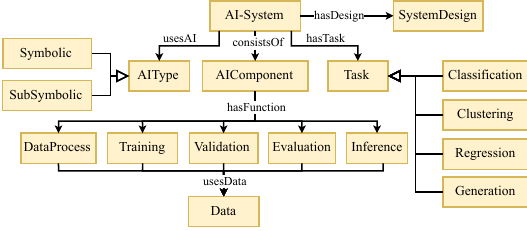}
    \caption{ISO 22989 ODP from a component and functions viewpoint. Classes and relations that are not mandatory for this viewpoint are not shown.}
    \label{fig:iso22989-aisystem}
\end{figure}

Within the first aspect, various types of AI (e.g., symbolic or subsymbolic), tasks (e.g., classification, clustering, regression, generation), system designs (e.g., cloud, edge, hybrid) their components, and functions (e.g., data processing, training, validation, evaluation, inference) are defined. A viewpoint of this first aspect is depicted in Figure \ref{fig:iso22989-aisystem}.

\begin{figure}[!h]
    \centering
    \includegraphics[]{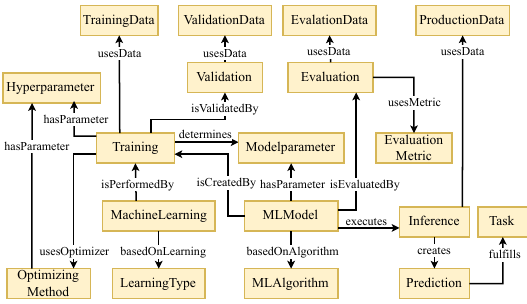}
    \caption{ISO 22989 ODP from an algorithmic viewpoint. Classes and relations that are not mandatory for this viewpoint are not shown.}
    \label{fig:iso22989-ai}
\end{figure}

The second aspect delineates the semantics and relationships of machine learning models, machine learning algorithms, and learning types in relation to the previously defined functions. Additionally, model parameters and hyperparameters are defined. A viewpoint of the second aspect is shown in Figure \ref{fig:iso22989-ai}.

\begin{figure}[!h]
    \centering
    \includegraphics[]{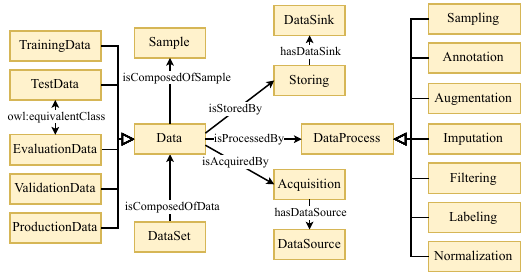}
    \caption{ISO 22989 ODP from a data viewpoint. Classes and relations that are not mandatory for this viewpoint are not shown.}
    \label{fig:iso22989-data}
\end{figure}

The third aspect describes the semantics and relationships between samples, data, and datasets. In addition to defining the data itself, the functions of storing data in a data sink and acquiring data from a data source are described. Furthermore, various types of datasets (e.g., training data, evaluation data, validation data, production data, test data) are defined. A viewpoint of this third aspect is illustrated in Figure \ref{fig:iso22989-data}. The third aspect serves to connect the first and second aspects within ISO 22989, particularly through the different types of data. 


\subsection{Description of the Alignment Ontology}
Up to this point, the ODPs have described the respective elements independently of each other. On the one hand, there are ODPs describing the technical processes, the technical systems, and the communication, such as VDI 3682, ISO 7498, and ECLASS. On the other hand, there are ODPs describing the AI systems, the AI components with their functions, and the data, such as ISO 22989. 

However, to create an information model capable of describing AI applications within automation systems, it is necessary to establish links between the two categories and their respective ODPs. To achieve this, an alignment ontology that connects the contents of multiple ODPs is needed. This alignment ontology constitutes the actual AIAS information model. It is created by importing all previously mentioned ontologies via \texttt{owl:import} statements. Through this approach, all imported ontologies remain in their own namespace, while the alignment ontology defines its own namespace, denoted as \texttt{AIAS:}.

\begin{figure*}[!b]
    \centering
    \includegraphics[]{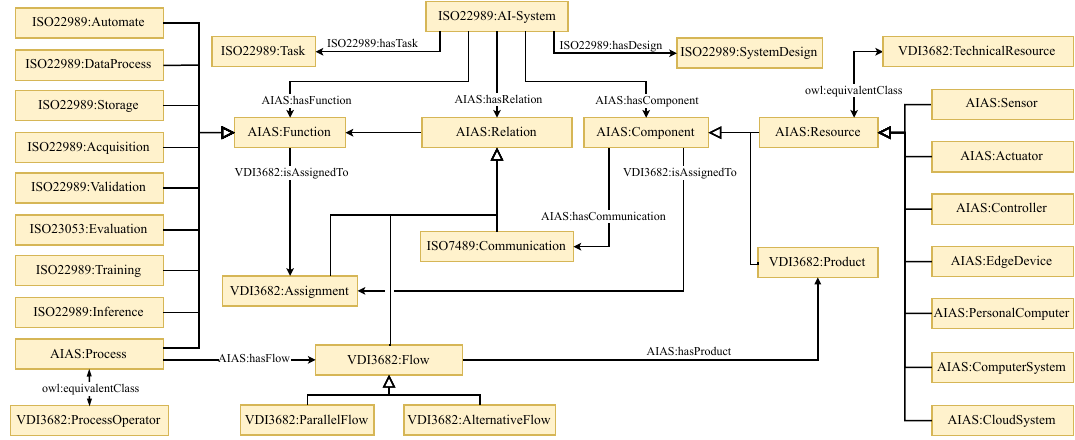}
    \caption{Core Concept of the alignment ontology AIAS, showing the connection of different standards through ODPs. Most of the concepts of the ODPs have been hidden to maintain readability.}
    \label{fig:aias-alignment-ontology}
\end{figure*}

The metamodel\footref{fn:metamodel} proposed by \citet{schieseck2023} forms the basic structure of the AIAS ontology, providing a framework for describing AI in automation systems. This metamodel includes the core classes \textit{function}, \textit{component}, and \textit{relation}. Subsequently, some of these core classes and some subclasses are successively replaced, extended or equated by the imported classes from the ODPs. This iterative process allows for the integration of specific domain knowledge and terms and ensures alignment with established standards. The core concepts of the AIAS alignment ontology and the underlying ODPs with their links are shown in a class diagram in Figure \ref{fig:aias-alignment-ontology}.

Following this approach, the AIAS ontology is initiated by creating the abstract core classes \texttt{AIAS:Function}, \texttt{AIAS:Relation}, and \texttt{AIAS:Component}. These classes lay the foundation for describing various aspects of AI within automation systems. Subsequently, to enable a high-level description of tasks and define the existing system design, the class \texttt{ISO22989:AI-System}, along with \texttt{ISO22989:Task} and \texttt{ISO22989:SystemDesign}, are imported into the AIAS ontology. 

The class \texttt{AIAS:Function} allows converting given inputs into outputs, thereby enabling  the achievement of the system goals. The possible types of functions within the system are described through the subclasses of \texttt{AIAS:Function}, which are imported from both VDI 3682 and ISO 22989. For instance, functions such as \texttt{ISO22989:Training}, \texttt{ISO22989:Inference}, and \texttt{ISO22989:Validation} are imported to describe various AI functionalities. Similarly, \texttt{VDI3682:ProcessOperator} is imported to describe certain technical process functionalities. To enhance clarity regarding the description of a technical process by the process operator, an alias \texttt{AIAS:Process} is created using the \texttt{owl:equivalentClass} statement, which simplifies and helps to understand the term process operator.

The class \texttt{AIAS:Component} allows describing the building blocks comprising the system. Two types of components are distinguished: resources and products. Resources are defined as system components that comprise the actual system and execute its functions. The definition of resources aligns with \texttt{VDI3682:TechnicalResource}, hence it is imported and equated with \texttt{AIAS:Resource}. Furthermore, the metamodel\footref{fn:metamodel} specifies seven types of resources: Sensor, Actuator, Controller, Edge Device, Personal Computer, Computer System, and Cloud System. These types can be equated with ECLASS or UNSPSC if necessary. In addition, products are defined as real-world entities that undergo transformation during a technical process. The definition of products is equivalent to \texttt{VDI3682:Product}.

The class \texttt{AIAS:Relation} allows the description of connections and relationships between system elements. Three types of relations are defined: assignment, communication, and flow. An assignment denotes a structural connection between functions and components, aimed at assigning a function to a component, which aligns with \texttt{VDI3682:Assignment}. Furthermore, communication signifies a connection between at least two components that allows the directed exchange of information between them, corresponding to \texttt{ISO7489:Communication}. Lastly, a flow describes the directed flow of a product through a technical process, equivalent to \texttt{VDI3682:Flow}.

Since the respective classes were imported from the ODPs, further relationships described in the ODPs are also integrated and imported into AIAS. For instance, relationships between training, model, inference, and data as shown in the Figures \ref{fig:iso22989-aisystem}, \ref{fig:iso22989-ai}, \ref{fig:iso22989-data} are incorporated, enabling a more detailed modeling of AI applications in automation systems.
\section{Application in an Industrial Scenario}  \label{sec:validation}
To demonstrate the use and to evaluate the presented AIAS information model, an exemplary industrial use case of an AI application was modeled. The use case was part of the research projekt EKI\footnote{\url{https://dtecbw.de/home/forschung/hsu/projekt-eki}\label{fn:eki}} and was specified and implemented with an industry partner. The modeling was done by instantiating and connecting the use case specific information. This way, a use case specific knowledge graph was created which can be found on GitHub\footref{fn:odps}, alongside a detailed documentation of its creation and further information.

\subsubsection*{Exemplary Use Case}
The use case revolves around a punching process that is carried out by a punching machine. The primary business objective was to minimize the maintenance costs associated with the stamping machine. 

The stamping machine is used to produce parts from blank metal sheets for the automobile industry. It comprises a support frame, a fixed lower die, and a movable upper die. The upper die is operated by an electric motor mounted on the support frame, with the driving force transmitted via a drive belt. The position of the upper die is monitored by a positional sensor. Over time, the wear of the drive belt leads to inaccurate control of the upper die, causing the parts produced to deviate from the strict tolerance requirements. Currently, the drive belt is replaced at regular intervals, resulting in high maintenance costs. Therefore, the primary business objective is to reduce maintenance costs by progressively classifying the condition of the drive belt.

Experiments by the industry partner indicate that a worn drive belt induces oscillations in the positioning of the upper die. Utilizing this insight, data recorded by sensors during the stamping process can be leveraged to infer the condition of the drive belt. 

The sensor data is transmitted to a controller via a bus network. The controller communicates with an edge device via Ethernet, while the edge device itself communicates with a cloud via internet. In the cloud, the model is trained and deployed for inference. A neuronal network is used as a model to analyze batches of the sensor data and to classify the belt condition.

\subsubsection*{Modeling}
The use case was successfully modeled using the AIAS ontology and the software \textit{Protégé}\footnote{\url{https://protege.stanford.edu/} \label{fn:protege}}. For instance, the communication between the edge device and the controller was modeled using an instance of the \texttt{ISO7489:Communication} class, connected with instances of the respective technical resources, \texttt{AIAS:EdgeDevice} and \texttt{AIAS:Controller}. Similarly, training in the cloud and deployment of the trained neuronal network were modeled using instances of the \texttt{VDI3682:Assignment} class, connected with instances of the functions \texttt{ISO22989:Training} and \texttt{ISO22989:Inference}, respectively. Throughout the modeling process, elements of the ISO 22989 ODP were used to provide further details, such as instances of the classes \texttt{ISO22989:Modelparameter},  \texttt{ISO22989:Hyperparameter} and \texttt{ISO22989:Data}. Additionally, the documentation of the task was made using an instance of \texttt{ISO22989:Classification}.

\subsubsection*{Additional Usage Options}
The presented type of information modeling enables querying, reasoning, and the application of rules, offering additional usage options.

Querying allows for retrieving specific information from a modeled use case, facilitating answers to questions such as those outlined in Tab. \ref{table:ComQuestions}. 
For example, to answer the question "\textit{Where was the model trained?}" one could utilize a SPARQL\footnote{\url{https://www.w3.org/TR/rdf-sparql-query/} \label{fn:sparql}} query like the one shown in Listing \ref{lst:SPARQL}.

\begin{lstlisting}[language=SPARQL, caption={Example SPARQL query to determine where the model was trained.} , basicstyle=\ttfamily\footnotesize, backgroundcolor=\color{really-light-gray}, label={lst:SPARQL}]
SELECT ?assignment ?component
WHERE
{
?training a ISO22989:Training .
?training AIAS:isAssignedTo ?assignment .
?component AIAS:isAssignedTo ?assignment . 
}
\end{lstlisting}

In addition to querying existing information, new information can be inferred using rules. Rules enable the formulation of more complex knowledge. For instance, within the AIAS information model, a rule could state: \textbf{IF} the inference is assigned to a cloud, \textbf{THEN} the AI system has the system design of a cloud system. Listing \ref{lst:SWRL} illustrates this rule using the Semantic Web Rule Language (SWRL\footnote{\url{https://www.w3.org/submissions/SWRL/} \label{fn:swrl}}).
\begin{lstlisting}[language=XML, caption={Example SWRL rule for reasonning that a given system has a cloud system design.} , basicstyle=\ttfamily\footnotesize, backgroundcolor=\color{really-light-gray}, label={lst:SWRL}]
AIAS:CloudSystem(?c) ^ VDI3682:Assignment(?a) ^ 
AIAS:isAssignedTo(?c, ?a) ^ ISO22989:Training(?t) ^ 
AIAS:isAssignedTo(?t, ?a) ->  ISO22989:hasDesign(AIAS:AISystem, AIAS:CloudDesign)
\end{lstlisting}

In some cases, it could be necessary to define constraints on the content and structure of the ontology and its individuals. Within the context of the AIAS information model, for example, it is beneficial to specify that every communication must occur between at least two components. Constraints of this nature can be established using the Shapes Constraint Language (SHACL\footnote{\url{https://www.w3.org/TR/shacl/} \label{fn:shacl}}). Listing \ref{lst:SHACL} demonstrates the implementation of the mentioned communication restriction using SHACL.
\begin{lstlisting}[language=XML, caption={Example SHACL shape for constraining communications.} , basicstyle=\ttfamily\footnotesize, backgroundcolor=\color{really-light-gray}, label={lst:SHACL}]
AIAS:Communication
    a sh:NodeShape  ;
    sh:targetClass ISO7489:Communication ;
    sh:property 
    [   sh:path AIAS:communicatesWith ;
        sh:minCount 2;  ].
\end{lstlisting}
\section{Discussion}
The AIAS enables the description of AI applications in automation systems by focusing on interdependencies between automation system components, AI components, and technical processes, as demonstrated through the exemplary modeling of the use case (R1).

The adoption of terms from standards such as ISO 22989, ISO 7489, or VDI 3682 ensures clear and unambiguous semantics (R2). Moreover, if necessary, the semantics of automation system components can be further enriched by importing ECLASS or UNSPSC into AIAS.

Utilizing ontologies in combination with OWL ensures the formalized capture of information. OWL's standardized representation language, maintained and defined by the W3C, makes it machine-readable and vendor-independent (R3).

The AIAS ontology, composed of several independent ODPs, offers extensibility and adaptability (R4). Additional ODPs can be easily imported and integrated into AIAS for more detailed descriptions. Updates to standards only require updating the relevant ODP, not the entire AIAS ontology.

In summary, it was shown that the AIAS information model fulfills the presented requirements (R1-R4) and enabled the modeling of an exemplary AI application for an automation system. Additionally, the querying based on SPARQL for easy retrieval of specific information was demonstrated. Furthermore, the formulation of complex knowledge through rules using SWRL and the specification of constraints using SHACL was shown.

Despite these advantages, there are some drawbacks to the approach. One drawback is the necessity of a detailed understanding of the AIAS ontology and its ODPs before modeling can even begin. Another drawback is the reliance on tools like Protégé, the use of which also requires expert knowledge. Even for proficient users of tools like Protégé, modeling can be tedious and time-consuming, as all individuals and relations for a particular use case have to be created manually. 
\section{Conclusion}
This paper introduces a formal model for describing Artificial Intelligence (AI) applications in automation systems. The concept allows the creation of a knowledge graph that encompasses the components of the automation system, the AI elements, the technical processes, and their interdependencies.

The formal model was implemented using ontologies, adopting a strategy of utilizing several smaller Ontology Design Patterns (ODPs) based on standards rather than constructing a single large and monolithic ontology. These smaller ODPs are combined to form an extendable and adaptable alignment ontology named AIAS. Specifically, an ODP based on the ISO 22989 standard is utilized to describe information about AI elements, while an ODP based on VDI 3682 is used for information about the automation system, technical processes, and components. Additionally, an ODP derived from ISO 7489 is created to depict communication within the automation system. The paper showcases the benefits of AIAS through an exemplary use case centered around a stamping process.

In summary, modeling with AIAS enables the formal description of AI applications for industrial use cases, facilitating knowledge acquisition through querying and knowledge generation via reasoning based on rules and constraints. Consequently, AIAS holds the potential to standardize the documentation and to streamline the integration, operation, and maintenance of AI applications within automation systems.
\section{Future Works}
In the future, we will add more subclasses to AIAS to broaden the scope of describing AI elements. This includes further subclasses of hyperparameters, model parameters, algorithms, and evaluation metrics. Additionally, we plan to create more rules and constraints to further improve the documentation process of AI applications. For instance, rules could be used to automatically classify the type of AI system based on the regulations outlined in the EU AI Act. Another potential of rules is to support the development process by suggesting suitable design patterns or best practices to developers.

Currently, querying was used to gather information from the knowledge graph of a single use case. However, as more developers adopt AIAS, we envision querying and comparing knowledge graphs from multiple use cases. This would enable suggesting previously solved and similar use cases to developers as they work on their projects. In addition to extending AIAS itself, we plan to develop a specialized AI modeling tool to streamline the modeling process and lower the barrier to entry for using AIAS. 
\section*{Acknowledgements}
This research is funded by \textit{dtec.bw – Digitalization and Technology Research Center of the Bundeswehr}. \textit{dtec.bw} is funded by the \textit{European Union – NextGenerationEU}
\begin{acronym}
  \acro{odp}[ODP]{Ontology Design Pattern}
  \acroplural{odp}[ODPs]{Ontology Design Patterns}
  \acro{owl}[OWL]{Web Ontology Language}
  \acro{aias}[AIAS]{Artificial Intelligence for Automation Systems}
  \acro{ai}[AI]{Artificial Intelligence}
\end{acronym}
\bibliographystyle{IEEEtranN}
\bibliography{references.bib}
\end{document}